\documentclass[aps, amsmath,amssymb,twocolumn,superscriptaddress,showpacs]{revtex4-1}

\usepackage{color}
\definecolor{green1}{rgb}{0,0.7,0.15}
\usepackage{verbatim}

\usepackage{bm}
\usepackage{graphicx}  
\usepackage{graphics}
\usepackage{epsfig}
\usepackage{soul}
\usepackage{cancel}
\usepackage{hyperref}
\usepackage{epstopdf}

\begin{document}

\newcommand{\alert}[1]{\textcolor{red}{#1}}
\newcommand{\dd}{\,{\rm d}}

\title{Fractal iso-contours of passive scalar in smooth random flows}

\author{M.~Vucelja}
\affiliation{Courant Institute of Mathematical Sciences, New York University, New York, NY 10012-1185, USA}
\affiliation{Kavli Institute for Theoretical Physics, University of California Santa Barbara, CA 93106}
\author{G.~Falkovich}
\affiliation{Department of Physics of Complex
Systems, Weizmann Institute  of Science, Rehovot, 76100 Israel}
\affiliation{Kavli Institute for Theoretical Physics, University of California Santa Barbara, CA 93106}
\author{K.~S.~Turitsyn}
\affiliation{Department of Mechanical Engineering, Massachusetts Institute of Technology, Cambridge, MA, 02139, USA}
\date{\today}



\begin{abstract}
We consider a passive scalar field under the action of pumping, diffusion and advection by a smooth flow with a Lagrangian chaos. We present theoretical arguments showing that scalar statistics is not conformal invariant and formulate new effective semi-analytic algorithm to model the scalar turbulence. We then carry massive numerics of passive scalar turbulence with the focus on the statistics of nodal lines. The distribution of contours over sizes and perimeters is shown to depend neither on the flow realization nor on the resolution (diffusion) scale $r_d$ for scales exceeding $r_d$. The scalar isolines are found fractal/smooth at the scales larger/smaller than the pumping scale $L$. We characterize the statistics of bending of a long isoline by the driving function of the L\"owner map, show that it behaves like diffusion with the diffusivity independent of resolution yet, most surprisingly, dependent on the velocity realization and the time of scalar evolution.
\end{abstract}
\maketitle

A fundamental problem in mixing is to describe  the iso-contours of the quantity which is mixed (called passive scalar in what follows). This is important for numerous practical applications. Renewed interest in this problem is related to the recent mathematical advances in describing random curves, particularly identifying an important class of curves SLE$_\kappa$ that can be mapped into a one-dimensional Brownian walk with the diffusivity $\kappa$. SLE stands for Schramm-L\"owner Evolution and presents a fascinating subject where physics meets geometry \cite{Schramm,GK,2005Cardy}. If lines belong to an SLE class, as was found, in particular, for isolines of vorticity and other quantities in turbulent inverse cascades  \cite{BBCF07etal,BBCFsqgetal,FM10}, conformal invariance provides exact formulae for the statistical description of the kind unimaginable before in turbulence studies.

It may seem that the simplest setting for mixing is when the velocity  is spatially smooth, which one encounters in many natural and industrial flows. An important problem of that type is large-scale mixing in the Earth atmosphere where the turbulence spectrum is approximately as steep as $k^{-3}$ \cite{NG};  the velocity gradients are then well-defined and the flow can be locally considered smooth. The same type of flows one encounters in the phase space of dynamical systems. Mixing in smooth flows is provided by exponential separation of trajectories and Lagrangian chaos. We consider an incompressible fluid flows which correspond to Hamiltonian flows in phase space. Generally, such flows have a positive
Lyapunov exponent $\lambda$ so they stretch any element into a long narrow strip. Additionally, our passive scalar is subject to random sources/sinks distributed in space with a correlation scale $L$. Resulting "passive scalar turbulence" in this (so-called Batchelor \cite{1959Batchelor}) regime has been extensively studied in terms of the one-point probability and multi-point correlation functions \cite{FGV}, one can even write down a closed expression for the probability of any given scalar field realization. Yet to the best of our knowledge, nothing is known about the properties of an infinite-point object, isoline. This may seem surprising since much is known about  isolines in a non-smooth (fully turbulent) velocity field \cite{Sreeni,DimotakisCatrakis,constantin1994geometric,BBCF07etal}. Apparently, the description of contours in the Batchelor regime is more involved. A conceptual difficulty is related to the lack of scale invariance. Indeed,  every long contour is simultaneously stretched to the scales far exceeding $L$ and contracted to the scales much less than $L$ in transverse directions. Technically, experimental and numerical
studies of passive scalar turbulence in the Batchelor regime are notoriously difficult because of a slow logarithmic convergence of the correlation functions with the resolution \cite{Steinberg}.

In this work we suggest a new, very effective, method of numerical simulation of the passive scalar turbulence in the Batchelor regime. The method is based on the analytic representation of the Lagrangian path integral. We carry out extensive numerical simulations of the passive scalar turbulence in two dimensions and find out that the isolines are non-fractal one-dimensional lines at the scales less than $L$. We then show that the isolines are fractal at the scales larger than $L$, and describe the statistics of the contour sizes and perimeters. Finally we explore relations of these isolines to SLE.

\begin{figure}
\includegraphics[width=3in,angle=0]{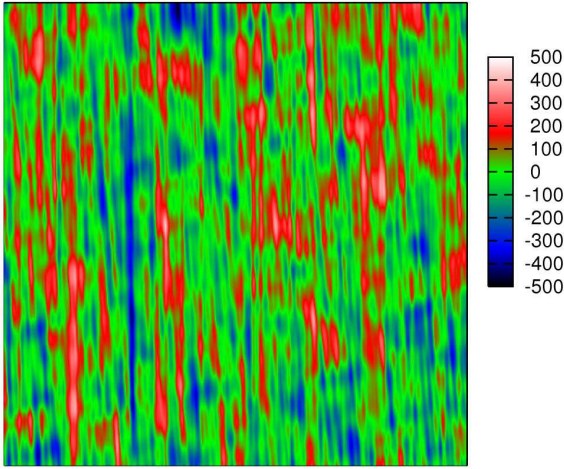}
\caption[Snapshot of the passive scalar field]{\label{scalar_snapshot}
A snapshot of a part of the passive scalar field of size $(75\, L)^{2}$, where $L$ is the forcing scale. The color bar shows the magnitude of $\theta$.}
\end{figure}

\begin{figure}
\includegraphics[width=3.41in,angle=0]{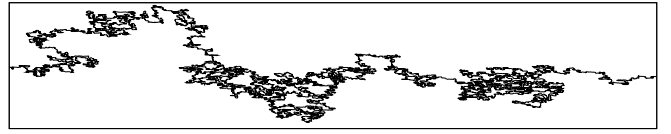}
\caption[Contour of passive scalar field]{\label{contour_snapshot}
A long scalar iso-contour in a window $2400 L \times 360 L$, where $L$ is the forcing scale. 
}
\end{figure}

Passive scalar $\theta$ is carried by the velocity  $\bm v$, forced by $\varphi$ and diffuses with molecular diffusivity $\kappa_{d}$:
\begin{align}
\partial _t \theta  + (\bm v \cdot \bm \nabla) \theta = \kappa_{d} \bm \nabla^{2} \theta + \varphi\,.\label{ad}
\end{align}
We take the pumping $\varphi(\bm r, t)$ to be white Gaussian with zero mean and variance $\overline {\varphi (\bm r , t ) \varphi(0, 0)} = \chi (r)\delta (t)$, where $\chi (r)$ decays faster than any power at $r > L$.
Our new computational method of generating the scalar field exploits  linearity of the advection-diffusion equation (\ref{ad}). To get a snapshot of $\theta({\bm r},t)$,  we sum over all of the blobs of scalar that hit the surface at random positions and random times in the past ($\nu$ times per unit time), see Figure~\ref{scalar_snapshot}. Each blob at its initial time $t_0$ has an isotropic shape centered around a random position $\bm r _c$ and of random amplitude $\Theta_{0}$: $\theta (t_0,{\bm r}_0) = \Theta_{0} \exp [- ({\bm r}_0-{\bm r}_c)^2/(2L^2)]$. The shape of such a blob at time $t$ is found by  characteristics:
\begin{align}
&\theta(t,{\bm r}_0) = \frac{\Theta_{0}L^2}{\sqrt{\det\hat I(t,t_0)}}e^{
-\frac{1}{2}({\bm r}(t)-{\bm r} _c)\hat I^{-1}(t,t_0)({\bm r}(t)-{\bm r} _c)} \,,
\end{align}
where $\bm r(t)\equiv\hat W(t,t_0) \bm r(t_0)$ defines the evolution operator $\hat W (t,t_0)$, obtained for a white in time and spatially linear velocity profile with
$$\left\langle{ \partial v_i(t)\over\partial x_j}{\partial v_k(t')\over\partial x_l}\right\rangle=D\delta(t'-t)[3\delta_{ij}\delta_{jl}-\delta_{ij}\delta_{kl}-\delta_{il}\delta_{kj}]\ .$$
The moment of inertia is $\hat I (t,t_{0}) \equiv \hat W(t,t_0)\hat W(t,t_0)^T$ $+\kappa_{d} \int ^t _0 \dd t' \hat W(t,t_0) \hat W(t',t_0)^{-1}[\hat W(t,t_0)\hat W^{-1}(t',t_0)]^T$. Seven values specify a blob at time $t$:  the symmetric matrix $\hat I(t,t_0)$, $\Theta_{0}$, ${\bm r} _c$ and $t_0$. The blob database is made large enough to ensure Gaussianity of $\theta$. To obtain the isolines of $\theta$ we used MATLAB's contourc function, an example of the contour can be seen in see Figure~\ref{contour_snapshot}. We measure the box-counting generalized fractal dimensions:
$ D_q \equiv \lim_{\varepsilon \rightarrow 0} \log(\sum^{N(\varepsilon)} _{i} p_i^q(\varepsilon))[(q-1)\log (\varepsilon)]^{-1}$. Here $p_{i}(\varepsilon)$ is the probability of finding a point in the $i-$th square of area $\varepsilon^{2}$, and $N(\varepsilon)$ is the number of $\varepsilon^{2}$ squares needed to cover the contour. For long contours, we study the statistics of bending by making the L\"owner map of the curve into a line \cite{Kennedy, BBCF07etal}. As one goes along the curve, the image moves along the line; if this motion (called driving function) is Brownian then the curve belongs to SLE.  We made several velocity realizations with different resolutions $r_{d}/L \in \{ 0.06, 0.03, 0.015\}$, pumping frequencies $\nu / \lambda \in \{ 0.05, 0.02, 0.01, 0.005 \}$ and scalar evolution times $T \lambda \in \{ 20, 50, 100\}$.

Let us now confront theoretical expectations with the results of numerics.
One expects that pumping alone would produce a Gaussian field $\theta$ whose zero isolines are smooth at the scales below $L$, while at larger scales they are equivalent to critical percolation called  SLE$_6$ (having the dimensionality $D_0=7/4$). Velocity field by itself does not change the statistics of  $\theta$ as it just rearranges it; the flow stretches isolines uniformly at the direction of the eigen-vector of the positive Lyapunov exponent and contracts them transversal to it. Non-trivial statistics of $\theta$ and its isolines arises from an interplay of velocity, pumping and finite diffusivity or finite resolution, which leads to the dissipation of
 $\theta$ and reconnection of isolines that came closer than the resolution scale $r_d$. We assume $r_d\ll L$.
At the scales between $L$ and  $r_d$, there is a cascade of passive scalar whose correlation functions of the scalar are logarithmic:  $\langle \theta ^n ({\bm 0}) \theta ^n ({\bm r}) \rangle \sim \ln ^n (L/r)$ \cite{1959Batchelor}.  Lower orders, $n<\ln(L/r)$,  correspond to Gaussian probability density function (PDF), while the PDF tails are exponential \cite{FGV}. The scalar field itself is thus non-smooth at $r<L$, what about its isolines? If the scalar was a Gaussian (free) field with logarithmic correlation functions, it would have the isoline with the fractal  dimensionality 3/2.
Let us show that the Gaussian PDF,
\begin{align}
{\cal P}\{\theta\} \propto \exp \left[ -  \frac{\mu}{2} \int \dd {\bm r} |\bm \nabla \theta (\bm r)|^2/\chi (\bm 0)\right]\,,
\end{align}
does not satisfy the respective Fokker-Planck equation:
\begin{align} \nonumber
\frac{\partial {\cal P}}{\partial t} =  &\frac{1}{2}\iint \dd \bm r_{1} \dd \bm r_{2} \frac{\delta ^2} {\delta
\theta (\bm r_{1}) \delta \theta (\bm r_{2})}\left[ K_{\alpha\beta}\nabla _\alpha \theta(\bm  r_{1}) \nabla
_\beta \theta (\bm r_{2}) \right.
\\
&\left.- \chi (\bm r_{1} - \bm r _{2}) \right] {\cal P} + \kappa_d \int \dd \bm r_{1} \frac{\delta}{\delta  \theta
(\bm r_{1})}\Delta \theta (\bm r_{1} ){\cal P}\!. \label{FP}
\end{align}
Indeed by substituting ${\cal P}$ in (\ref{FP}) we see that already the first term gives a non-vanishing contribution (highest order in $\theta$)
\begin{align} \nonumber
&\iint \!\! \dd \bm r_{1} \dd \bm r_{2} D {\cal P} \{ 2[\bm \nabla \theta (\bm r_{1}) \cdot \bm \nabla \theta (\bm r_{2})]^2
\\
&- |\bm \nabla
\theta (\bm r_{1})|^2 |\bm \nabla \theta (\bm r_{2})|^2 \}\,.
 \end{align}
Indeed, we know that correlation functions of $\theta$ include cumulants  \cite{BCKL}. 
One may argue that
the cumulants contain
less logarithmic factors than reducible terms and are small \cite{FGV}. However, the properties
of isolines must depend on those cumulants, since they contribute
the correlation functions of the gradients in the main order. It is straightforward to
establish that the correlation functions of the passive scalar are not conformal invariant, i.e., for instance,
the four-point function does not have the form
\begin{align} \label{four-point}
&F_4 = f\left(\frac{r_{12}r_{34}}{r_{13}r_{24}},\frac{r_{12}r_{34}}{r_{14}r_{23}}
\right) (r_{12}r_{34}r_{14}r_{23}r_{13}r_{24})^a\,.
\end{align}
Indeed, we know from \cite{BCKL} that $F_4 = F({\bm r}_{12}, {\bm
r}_{34})+F({\bm r}_{13}, {\bm r}_{24})+F({\bm r}_{14}, {\bm r}_{23})$ which is
compatible with (\ref{four-point}) only for Gaussian statistics, which is not the case, as we have shown. Therefore, passive scalar is not in any way close to a free field.

Note in passing that one can also show that the scalar statistics is not conformal invariant in a compressible 1d Batchelor-Kraichnan model, where velocity is Gaussian with the zero mean and the variance
$\langle v (x,t)v
(0,0)\rangle=\delta(t)\Bigl[K_0-x^2\Bigr]$.
Correlation function  of the scalar satisfy $(\partial_t+\sum
x_{ij}^2\partial_i\partial_j)\langle \theta_1 \theta_2 \rangle=\chi(x_{12})$.
At $x\gg L$ the pair correlation function of the scalar has parts linearly
growing with time while the structure functions are finite:
$\langle(\theta_1-\theta_2)^2\rangle\approx \chi(0)\ln(x_{12}/L)$ \cite{CKV}.
The two-point correlation function of the gradients is
$\langle\omega_1\omega_2\rangle\approx \chi(0)x_{12}^{-2}$. The four-point
correlation function of the gradients can be calculated exactly for
$\chi(x)\propto\exp(-x/L)$ using the method of \cite{CKV}:
$\langle\omega_1\omega_2\omega_3\omega_4\rangle=
\chi^2(0)\sum
 [x_{mn}^{-2}x_{kl}^{-2}+2(x_{mn}+x_{kl})^{-2}/(x_{mn}x_{kl}) ]$ for $m\not=n\not=k\not=l$. Both parts (Gaussian and cumulant) are
zero modes of the operator ${\cal L}=\sum\partial_i\partial_jx_{ij}^2$. The
cumulant is not conformal invariant: under the transformation of coordinates
$x'=(ax+b)/cx+d)$ with $ad-bc=1$ it changes its form rather than just acquire
the factors (Jacobians) $(dx'/dx)^{\Delta}=1/(cx+d)^{2\Delta}$. The dimension
of $\omega$ is $\Delta=1$ as can be seen from the invariance of the pair
correlation function:
$(x_{1}-x_2)^{-2}(cx_1+d)^{2\Delta}(cx_2+d)^{2\Delta}=(x_1'-x_2')^{-2}=
(x_{1}-x_2)^{-2}(cx_1+d)^{2}(cx_2+d)^{2}$. We thus conclude that the scalar
field is not conformal invariant.

\begin{figure}
\includegraphics[width=3.42in]{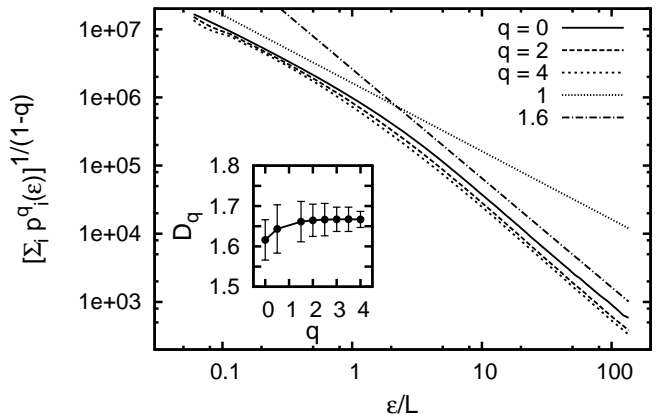}
\caption[]{\label{fig:average_Dq_nu0.01.eps} The generalized box counting fractal dimension $D_{q}$ is scale dependent. Below the forcing scale $L$ the curves are smooth with $D_{q} = 1$, while above $L$, the contour seems to be a simple fractal with $D_{q}$ in between $1.55 \div 1.7$. The pumping frequency was $\nu/ \lambda = 0.01$.}
\label{fig:dim}
\end{figure}

If one tries to find an analogy with a non-Gaussian field having logarithmic correlation functions, such is  the
height function built on independently oriented loops from the $O[n]$ model \cite{CZ}, deviations of $|\kappa-4|$ and $|D-3/2|$ are proportional to cumulants in this case. Another much exploited similarity is between the passive scalar and the vorticity cascade of two-dimensional turbulence; vorticity isolines was shown to have multifractal isolines with dimensionalities changing from $3/2$ to $1$ \cite{BBCFsqgetal}. Despite all these suggestive similarities, let us show that, contrary to these expectations, scalar isolines are smooth below $L$.
Figure~\ref{fig:average_Dq_nu0.01.eps} shows the box-counting dimensionalities of  contours. We see that, contrary to the expectations, the scalar isolines are smooth below $L$, which is quite natural from the physical perspective since all the factors (velocity, pumping, diffusion) are smooth at these scales. In particular, that means that the scalar field non-smoothness is related to the discontinuities of  $\theta({\bf r})$ across (smooth) isolines.
\begin{figure}
\includegraphics[width=3.42in]{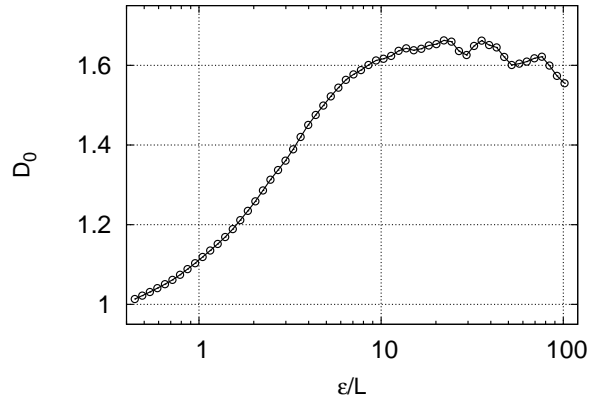}
\caption[]{\label{fig:ave_D0_fig.eps} Box counting fractal dimension
 $D_{0}$, for $\nu/ \lambda = 0.01$, estimated from the derivative $\dd\log N(\varepsilon)/\dd \log(L/\varepsilon)$.
It is scale dependent and we see that bellow the forcing scale
 $L$ the curves are smooth with $D_0 = 1$, while above $L$, the contour seems to be a
 fractal with $D_0$ in between $1.55 \div 1.7$.}
\end{figure}

\begin{figure}
\includegraphics[width=3.42in]{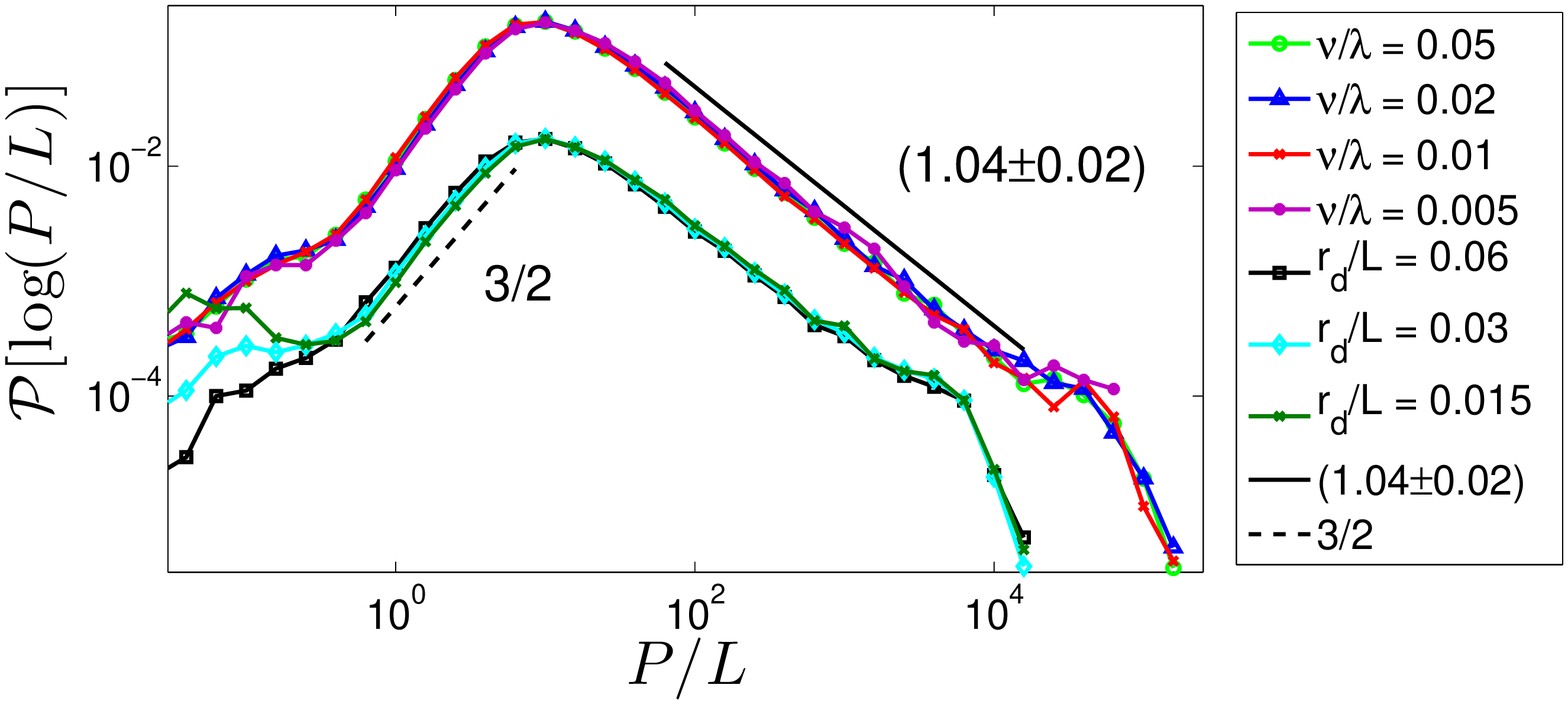}
\caption[PDF of $\log(P/L)$]{\label{fig:lnP_hist.eps}
The PDF of perimeters for different pumping frequencies (upper three curves) and
resolutions (lower three curves shifted down by dividing by 10).}
\end{figure}

\begin{figure}\includegraphics[width=3.42in]{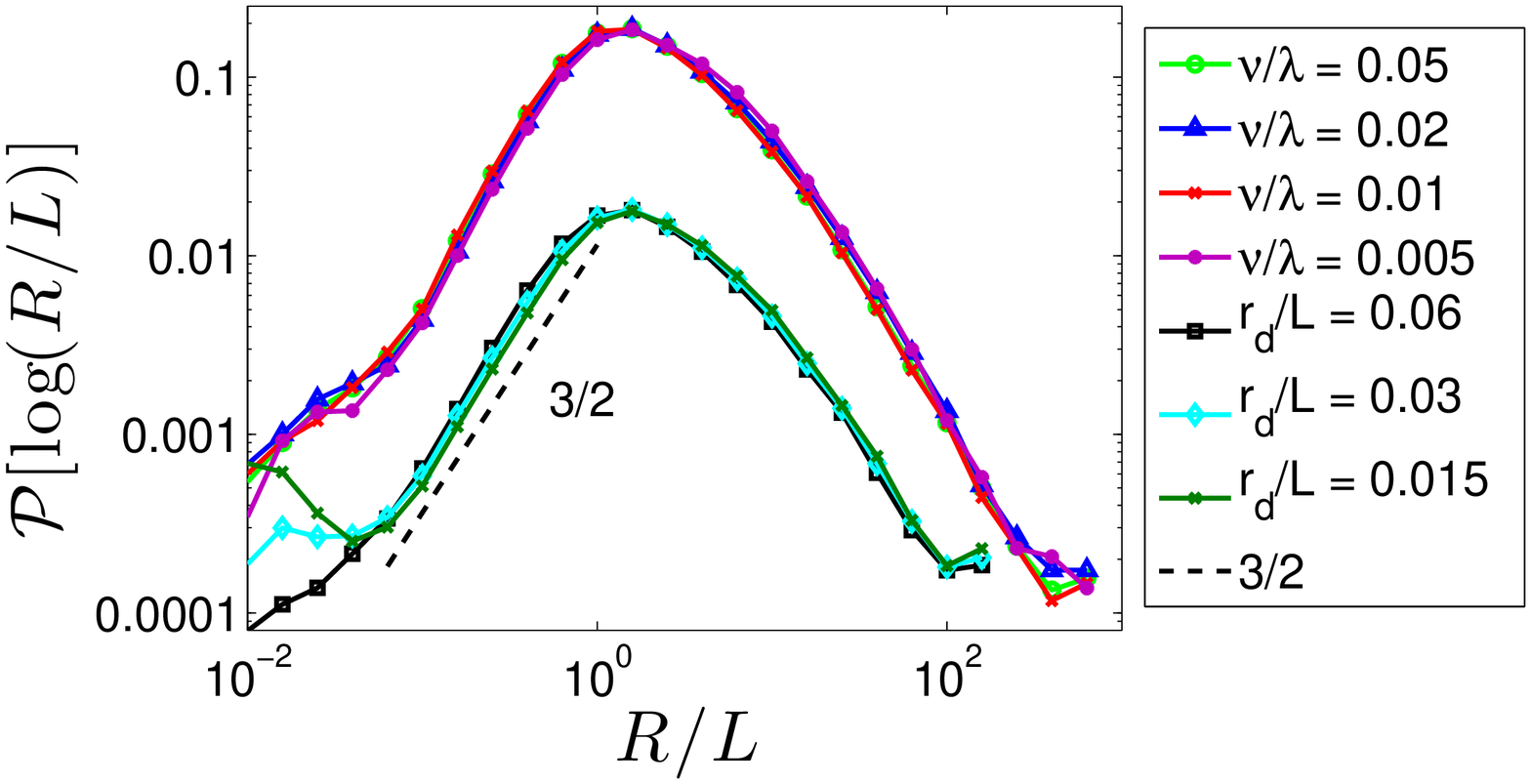}
\caption[PDF of $\log(R/L)$]{\label{fig:lnR_hist.eps}
The PDF of sizes for different pumping frequencies (upper three curves) and
resolutions (lower three curves shifted down by dividing by 10).}
\end{figure}

Let us now discuss the probability density function (PDF) of contour perimeters $P$ and sizes characterized by the mean radius $R \equiv \sqrt{\langle (\bm \rho - \langle \bm \rho \rangle)^{2} \rangle}$, here $\bm \rho$ denotes the pairs of points parameterizing a contour and averaging $\langle \cdot \rangle$ is done over the points. Figures~\ref{fig:lnP_hist.eps},\ref{fig:lnR_hist.eps} present the PDFs of $\log (P/L)$ and $\log(R/L)$ and show that they depend neither on the resolution for $r_{d}/L = 0.015 \div 0.06$ nor on the pumping frequency for $\nu /\lambda = 0.005 \div 0.05$. In both figures, the upper three curves are for different $\nu /\lambda$ and the same  $r_{d}/L = 0.06$; the lower three curves (PDF of those was divided by 10) differ in $r_{d}/L$, while $\nu /\lambda  = 0.01$. The only difference one can distinguish in the lower curves in Figures~\ref{fig:lnP_hist.eps},\ref{fig:lnR_hist.eps} is the appearance of the secondary maximum at small scales. This shows an abundance of diffusion-scale contours since the maximum appears only for the best resolution with the scale $0.015L$ comparable to the molecular diffusion scale, which is $\sqrt{\kappa_{d}/\lambda} = 0.01L$ in all runs. PDFs do not seem to depend on the velocity realization so Figures~\ref{fig:lnP_hist.eps},\ref{fig:lnR_hist.eps} were obtained by averaging over different realizations.

Let us consider separately left and right tails of the PDFs.
Since the probability of contours much larger than $r_d$ is independent of diffusion then it is determined by an interplay of stretching and pumping. Figures~\ref{fig:lnP_hist.eps},\ref{fig:lnR_hist.eps} show that the left tails of both ${\cal P}(P)$ and  ${\cal P}(R)$ look like a power law with the power $3/2$. Contours shorter than $L$ must appear when pumping cuts a piece off a thin long contour, the probability of such a cut is   $\propto P\propto R$ (small contours are smooth so that  $P\propto R$). Extra factor $\sqrt{P}\propto \sqrt{R}$ in the PDF may appear because to be observed small contours need to survive without being swallowed by further pumping events, the lifetime is likely to be  $\propto \sqrt{P}\propto \sqrt{R}$. Since creation and survival are independent events, their probabilities are multiplied.

What one may expect for the PDFs of contour sizes at scales exceeding $L$? It is tempting to assume that a long contour appears due to an evolution undisturbed by pumping during a long time $t$, the length of such contour is $L\exp(\lambda t)$ as long as it does not exceed $L^2/r_d$. The probability that pumping did not act during the time $t$ is given by a Poisson law $\exp(-\nu t)$ where $\nu$ is the pumping correlation time (in our algorithm, an average time between producing blobs of $\theta$ in a given place). We then obtain
${\cal P}(R)=\int dt \exp(-\nu t)\delta\bigl(R-\exp(\lambda t)\bigr)\propto R^{-1-\nu/\lambda}\,,$ which would mean that the PDF tail is non-universal and depends on the statistics of pumping and velocity. If that was true, the same tail one would expect for the PDF of perimeters as well. However, the above consideration totally disregards the fractal nature of long contours (shown in Figure~\ref{fig:dim}). We have checked that the fractal dimensions within our error bars were the same for different pumping frequencies $\nu$, see Table~\ref{tab:nudiff}. In line with long contour fractality, the right tails of the PDFs of $P$ and $R$ are very different. The tail of the PDF of sizes, ${\cal P}(R)$, looks log-normal, see Figure~\ref{fig:lnR_hist.eps}. The tail of ${\cal P}(P)$ looks like a universal power law, independent of the resolution and the pumping frequency $\nu$, see  Figures~\ref{fig:lnP_hist.eps}. In log coordinates the tail is close to $1/P$ so it may be that  ${\cal P}(P)\propto P^{-2}$,
yet we were unable to derive this law theoretically so far.
\begin{table}[h]
\begin{tabular}{|c||c|c|c|}
\hline
$\nu/\lambda$& $D_{0}$&$D_{2}$ &$D_{4}$ \\
\hline
0.05 &$1.61\pm0.05$&$1.66\pm0.02$&$1.67\pm0.02$  \\
0.02 & $1.62\pm0.03$& $1.65\pm 0.05$ & $1.65 \pm 0.05$  \\
0.01 & $1.62\pm0.05$ & $1.67\pm0.05$ & $1.67\pm0.03$  \\
0.005 &$1.62\pm0.03$&$1.68\pm0.06$ & $1.68\pm0.04$ \\
\hline
\end{tabular}
\caption{\label{tab:nudiff}Generalized fractal dimensions, for different pumping frequencies $\nu/\lambda$.}
\end{table}

Within our accuracy, we cannot see any difference between the dimensionalities of the different orders and conclude that our contours are mono-fractals in distinction from multi-fractal iso-vorticity contours in a direct cascade of $2d$ turbulence \cite{BBCF07etal}. This difference might be due to the fact that all parts of our scalar contours go through the same history of velocity, while parts of a long vorticity contour may  have  different histories.

Looking at Figures~\ref{fig:lnP_hist.eps},\ref{fig:lnR_hist.eps}, a natural question is whether the positions of the maxima depend on the resolution. Pumping produces more or less circular contours of the radius $L$, which are then deformed into ellipsoids of increasing eccentricity by the velocity field. Pumping continues to act by bending elongated contours. Those contours that on average have not changed much by this bending disappear after reaching the length of order $L^2/r_d$ and respectively the width of order the coarse-graining scale $r_d$. One then asks if the scale $L^2/r_d$  is special, apart from $L$ and $r_d$. Numerics give a negative answer: lower curves in Figures~\ref{fig:lnP_hist.eps},\ref{fig:lnR_hist.eps} compare  runs with three different $r_d$ and show that the PDFs do not depend on the resolution for the scales exceeding $r_d$.
 In particular, the PDFs of the sizes  have the maximum at the pumping scale $L$ independently of the resolution (or diffusion scale). We checked that the statistics is practically the same for either an average distance of the contour points from their center of mass or the maximal distance between the points of the contour (gyration radius). The PDFs of perimeter also have maxima independent of the diffusion scale (at the size approximately $2\pi L$).
 The only difference one can distinguish in the lower curves in Figures~\ref{fig:lnP_hist.eps},\ref{fig:lnR_hist.eps} is the appearance of the secondary maximum at small scales for the curve with the best resolution, there the resolution scale $0.015L$ is comparable with the diffusion scale, which is $0.01L$ in all runs; in other words, there is an abundance of diffusion-scale contours.
\begin{figure}
\includegraphics[width=3.41in]{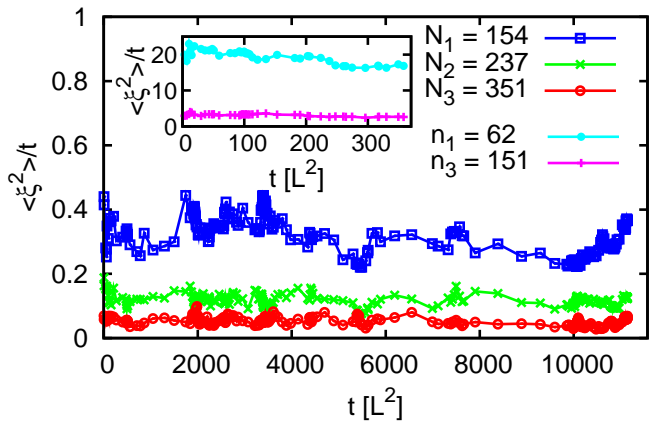}
\caption[]{\label{fig:ave_xi2_c1_elonger.eps} Effective diffusivity of the driving function $\xi(t)$ for velocity realizations ($1\div3$), where $t$ is time in L\" owner's eq., see \cite{Schramm}. The curves are ensemble averages of $N_{i}$ contours (inset: of $n_{i}$ contours) from the $i-$th velocity realization. 
Inset shows the effective diffusivity for velocity realizations 1 and 3 after contraction by $L/r_{d}$.}
\end{figure}

Quite unusual statistics of the passive scalar lacking scale-invariance has been found at $r>L$: multi-point correlation functions strongly depend on geometry \cite{Largeetal}. This is because of the highly anisotropic "strip" structure of the field seen clearly in Figure~\ref{scalar_snapshot}. Let us discuss the statistics of bending for the isolines extending for such long distances. If there was only pumping then on the scales larger than $L$ the scalar would be a short-correlated field whose nodal lines are equivalent to critical percolation i.e. SLE$_6$. Without diffusion and with infinite resolution, velocity only distorts the field. Of course, distorted field is not SLE \cite{Kennedy}; for instance, stretching vertically a chordal SLE in a half-plane one adds to the driving function extra intervals of no change, that diminishes $\kappa$ and provides for a finite correlation scale (equivalently, finite correlations in L\"owner time $t$ which is the coordinate along the contour). And yet deforming it back (with a time-dependent distortion factor $\exp(2\lambda T)$) we would get the same SLE$_6$. While the restoration procedure itself may be not very practical since real flows consist of many such domains oriented randomly, the very possibility of it means potential availability of very useful exact formulae describing the statistics of contours. For example, one may be able to describe crossing and surrounding probabilities (like Cardy-Smirnov formula \cite{92Cardy,Smirnov}) with just a simple re-scaling. However, a finite resolution/diffusion leads to irreversible changes (and makes the statistics stationary). Indeed, when velocity distorts the contours, it causes some distances in the contracting directions to become less than the resolution scale $r_d$ which leads to reconnections and disappearance of thin contours. Now it is highly nontrivial if any trace of initial SLE can be recovered, and if yes, after what contraction. One may try to estimate the stretching factor empirically by measuring the aspect ratio of the boxes one can fit the contours into; we found out that this approach does not work since such an aspect ratio fluctuates strongly (by several orders of magnitude).

Let us remind that Oseledec theorem and a general theory of Lagrangian chaos with diffusion \cite{FGV} state that any finite-point statistics is independent of the velocity realization and can be understood under the assumption that the effective contraction factor is $L/r_d$. This is based on the fact that Lyapunov exponents are self-averaging and the mean lifetime of scalar blobs is  $\lambda^{-1}\ln(L/r_d)$. If one is to extend this reasoning to the statistics of an infinite-point object, then one expects the isoline statistics to depend strongly on the resolution scale $r_d$ and be independent of the velocity realization in a steady state and for sufficiently long contours that appear after a long history of stretching.
As we now see, both predictions fail spectacularly.
\begin{figure}
\includegraphics[width=3.41in]{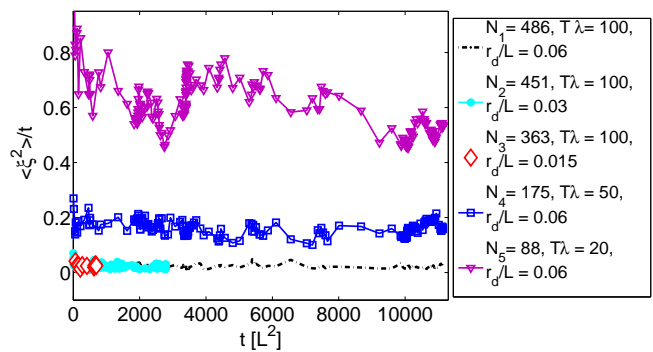}
\caption[]{\label{fig:kappa_diffresol_v02.eps} Diffusivity $\kappa$ for different resolution scales and the evolution times $T \lambda$ for the same velocity realization.  The values $N_{i}$ show the number of contours we averaged over and label the curves. Graphs $N_{1}$,$N_{2}$ and $N_{5}$ show that $\kappa$  depends on the  evolution time $T\lambda \in \{20,50,100\}$, while $N_{1}$, $N_{2}$ and $N_{3}$ show that $\kappa$ does not depend on the resolution $r_{d}/L\in \{0.015,0.03,0.06\}$. The curves of different resolutions are of different length since the time of the L\" owner map is measured in the units of length squared and for a better resolution we used a smaller physical window (all scalar snapshots had $40000^2$ pixels, but in terms of the forcing scale they had $(600 L)^2,(1200 L)^2,(2400 L)^2$ respectively).}
\end{figure}

It was already stated above - we characterize the behavior of a single line  by the driving function of the L\"owner map.   Figure~\ref{fig:ave_xi2_c1_elonger.eps} shows that driving functions behave roughly as that of diffusion and so can be characterized by the diffusivity $\kappa$. It is independent of the resolution scale (this may be related to the problem of noise sensitivity in the statistics of random curves \cite{BSK}), as seen in Figure~\ref{fig:kappa_diffresol_v02.eps}.  It is also independent of the contour size as long as it is larger than $L$, yet $\kappa$ is different in different velocity realizations even on the same very long L\" owner timescale (i.e. for very long contours). Most dramatically, Figure~\ref{fig:kappa_diffresol_v02.eps} shows that $\kappa$ depends strongly on time $T$ (of stretching) on extremely long timescales far exceeding the time  $\lambda^{-1}\ln(L/r_d)$ during which the scalar field itself acquires stationary statistics. We conclude that long contours undergo stretching and their statistics changes on a much larger timescale than the scalar blob lifetime. Let us stress the discrepancy: $\kappa$ depends strongly on the distortion and yet it is independent of $r_d$ (at least within the limits we studied), that is the effective distortion factor for long isolines is not $L/r_d$ (as it is for multi-point statistics). On the other hand, note that choosing the contraction factor equal to $L/r_d$ one obtains $\kappa$ of the "right" order of magnitude (between 3 and 17 for different velocity realizations, see inset of Figure ~\ref{fig:ave_xi2_c1_elonger.eps}). One is tempted to explore whether one can recover SLE contours (despite all the loss of information due to reconnections) by fine-tuning the distortion factor, possibly by requiring either restoration of statistical isotropy or shortest correlation time of $\xi(t)$.
 Further studies with extensive statistics are needed to sort out which properties of the critical percolation are retained by the large-scale statistics of passive scalar contours in the Batchelor regime.

This research was supported  by the NSF grant PHY05-51164 at KITP, and by the grants of BSF, ISF and Minerva foundation  at the Weizmann Institute. We benefitted from discussions with I. Binder, G. Boffetta, D. Dolgopyat, A. Celani and K. Khanin.

\bibliography{Bib/VuceljaBib}
\end{document}